# A SURFACE-RENEWAL MODEL FOR CONSTANT FLUX CROSS-FLOW MICROFILTRATION

Shaopeng Jiang[1] and Siddharth G. Chatterjee[2]

[1] Department of Civil and Environmental Engineering, 151 Link Hall
Syracuse University, Syracuse, New York 13244, USA

[2] Department of Paper and Bioprocess Engineering, SUNY College of Environmental Science and Forestry, 1 Forestry Drive, Syracuse, New York 13210, USA

Correspondence to: Siddharth G. Chatterjee (E-mail: *schatterjee@esf.edu*)

**ABSTRACT**[1]

A mathematical model using classical cake-filtration theory and the surface-renewal concept is formulated for describing constant flux, cross-flow microfiltration (CFMF). The model provides explicit analytical expressions for the transmembrane pressure drop (TMP) and cake-mass buildup on the membrane surface as a function of filtration time. The basic parameters of the model are the membrane resistance, specific cake resistance, and rate of surface renewal. The surface-renewal model has two forms: the complete model, which holds for compressible cakes, and a subsidiary model for incompressible cakes, which can be derived from the complete model.  The subsidiary model is correlated against some of the experimental TMP data reported by Miller *et al*.

---

(2014) for constant flux CFMF of a soybean-oil emulsion in a cross-flow filtration cell having unmodified and surface-modified, fouling-resistant membranes, and has an average root-mean-square (RMS) error of 6.2%. The complete model is fitted to the experimental TMP data reported by Ho and Zydney (2002) for constant flux microfiltration of a bovine serum albumin solution in a stirred cell using polycarbonate track-etched membranes and has an average RMS error of 11.2%. This model is also correlated against the TMP data of Kovalsky *et al*. (2009) for constant flux yeast filtration in a stirred cell (average RMS error = 9.2%).

**KEYWORDS:** Constant flux, Cross flow, Membrane, Microfiltration, Surface renewal

**INTRODUCTION**

Cross-flow membrane filtration technology has come into wide use in the chemical and biotech industries globally, and is also becoming common in wastewater treatment. In cross-flow filtration, an incoming feed solution or suspension passes over the surface of a membrane with the permeate flow being that portion of the liquid that passes through the membrane in a direction perpendicular to the direction of the main flow. The permeate flux depends upon the membrane characteristics, fluid velocity, viscosity, dissolved/suspended solids concentration, transmembrane pressure drop, temperature, and membrane fouling. For a constant transmembrane pressure drop (TMP), the permeation flux declines with the progress of process time due to fouling of the membrane by pore blocking, concentration polarization, and cake buildup on its surface.

The surface-renewal concept has been used to theoretically model constant pressure, cross-flow microfiltration and ultrafiltration by a number of workers.[1-9] Compared to the film and boundary-layer models of cross-flow membrane filtration, the surface-renewal model has the potential to more realistically describe the transfer of dissolved/suspended solids due to random hydrodynamic impulses generated at the membrane-liquid interface, e.g., due to membrane roughness or by the use of spacers or turbulence promoters.

The majority of work reported in the literature on cross-flow membrane filtration is for constant TMP operation with only a few studies being available for constant flux conditions. Two examples of the latter are the papers of Ho and Zydney[10] and Kovalsky *et al*.[11] who presented numerical models for CFMF under constant flux conditions. However, these models do not explicitly include the effect of flow instability, generated by the axial flow of liquid over the membrane surface, on membrane performance (i.e., TMP), although the combined pore-blockage, cake-filtration model of Ho and Zydney[10] contains a parameter *f* (not to be confused with the age-distribution function *f* of surface elements to be discussed later), which, according to these authors, "can also account for the reduction in protein deposition due to any back-flux phenomena, including the effects of crossflow and any long-range electrostatic interactions."

Since the constant flux mode of operation is becoming increasingly more common, in the present work, a mathematical model for constant flux CFMF is developed that uses the surface-renewal concept through which the effect of flow instability on membrane performance is taken into account. The surface-renewal model presented herein, which provides explicit analytical expressions for the TMP and cake-mass buildup on the membrane surface as a function of filtration or process time (unlike the models of Ho and Zydney[10] and Kovalsky *et al.*[11]), has two versions: the complete model, which includes the effects of cake compressibility, and a subsidiary model valid for incompressible cakes, which can be derived from the complete model. The subsidiary model is correlated against some of the experimental TMP versus process time data reported recently by Miller *et al.*[12] for constant flux CFMF of a soybean-oil emulsion in a cross-flow cell having unmodified and surface-modified, fouling-resistant membranes. The complete model is fitted to the experimental TMP data of Ho and Zydney[10] for constant flux microfiltration of a bovine serum albumin (BSA) solution in a stirred cell using polycarbonate, track-etched membranes and also to the TMP data of Kovalsky *et al.*[11] for constant flux yeast filtration in a stirred cell.

**SURFACE-RENEWAL MODEL**

In the surface-renewal model (see Hasan *et al.*[8]), it is postulated that the dominant fouling mechanism responsible for permeate-flux decline is cake formation with pore blocking occurring only during the initial moments of filtration and which effect, if important, can be incorporated into the membrane resistance $R_m$. Flow instabilities are assumed to constantly bring fresh liquid elements from the bulk liquid to the membrane-liquid interface. A liquid element resides at the membrane surface for a certain amount of time after which it departs and re-mixes with the bulk liquid. Above the surface elements, the liquid is assumed to be well mixed and where the concentration of solids is held constant due to a high rate of transport (because of flow instability) from this location to the bulk liquid. Gradually, a cake layer builds up on the membrane wall which causes an increase in the TMP with process time under constant flux conditions. In order to model this process, which is the chief objective of this paper, it is assumed that during the residence time $t$ of a liquid element at the membrane surface, TMP buildup within it can be described by classical cake-filtration theory.[13] From this theory, the pressure drop $\Delta p_c$ across the cake in a surface element with a residence time of $t$ can be expressed as:[13]

$$(\Delta p_c)^{1-n} = K_r t \qquad (1)$$

where

$$K_r = \mu J^2 c_b \alpha_0 \qquad (2)$$

In the above, $J$ = constant permeate flux, $\mu$ = viscosity of the filtrate, and $c_b$ = mass of solids deposited on the membrane surface per unit volume of filtrate passing through it (approximately equal to the bulk or feed concentration of solids), which is assumed to be constant. The two compressibility parameters of the cake $\alpha_0$ and $n$ are empirical constants with $n$ being the compressibility coefficient of the cake, which is zero for incompressible sludges, and lies between 0.2 and 0.8 for compressible ones.[13] They are related to the specific cake resistance $\alpha$ through the empirical equation:

$$\alpha = \alpha_0 \Delta p^n \qquad (3)$$

where $\Delta p$ is the (total) TMP in a surface element that is given by:

$$\Delta p = \Delta p_c + \Delta p_m \qquad (4)$$

Here, $\Delta p_m$ is the pressure drop across the membrane, which can be expressed as:

$$\Delta p_m = \mu J R_m \tag{5}$$

where, as mentioned earlier, $R_m$ is the membrane resistance.

The mass $m_c$ of solids accumulated in the liquid element per unit area of the membrane surface during the time period of $t$ is given by:

$$m_c(t) = J c_b t \tag{6}$$

The surface of the membrane at a given value of the process time $t_p$ during the filtration is visualized as being populated by a mosaic of liquid elements that have ages in the time interval of zero to $t_p$. If we denote the age-distribution or residence-time distribution (i.e., RTD) function of the surface elements by $f(t, t_p)$, the age-averaged cake pressure drop $\Delta p_{c,a}$ and age-averaged cake mass $m_{c,a}$ accumulated per unit area of the membrane surface at process time $t_p$ may be written as:

$$\Delta p_{c,a}(t_p) = \int_0^{t_p} \Delta p_c(t) f(t, t_p) dt \tag{7}$$

and

$$m_{c,a}(t_p) = \int_0^{t_p} m_c(t) f(t, t_p) dt \tag{8}$$

In Eq. (7), the cake pressure drop $\Delta p_c$ in an individual surface element is treated as an 'information content' or 'stress level' of the element. Thus, at given values of the process time $t_p$ and the imposed, constant permeate flux $J$, an older element will require a greater $\Delta p_c$ compared to a younger element as can be seen from Eq. (1).

As demonstrated by Zhang and Chatterjee,[9] using different speculative hypotheses about the behavior of liquid elements on the membrane surface, which correspond to different startup conditions, different RTD functions [i.e., $f(t, t_p)$] can be derived. These can then be used in Eqs. (7) and (8) to develop expressions for the age-averaged pressure drop across the cake and cake mass buildup. In this work, the Danckwerts distribution function[14] will be used to represent the ages of surface elements,[2,8,9] i.e.,

$$f(t, t_p) = \frac{S e^{-St}}{1 - e^{-S t_p}} \tag{9}$$

where $S$ (assumed to be constant) is the rate of renewal of liquid elements at the membrane surface and is a hydrodynamic parameter. It increases with velocity of the main flow[1-3] and can also be looked upon as a "scouring" term that represents the removal of deposited material from the membrane wall,[5] which will depend upon the level of flow instability. From dimensional considerations, Hasan et al.[8] have proposed a correlation for $S$ as a function of the diameter of the membrane channel, axial flow velocity, relative roughness of the membrane wall, and viscosity and density of the feed suspension.

**Complete Model: Compressible Cake ($n \neq 0$)**

From Eqs. (1) – (9) it can be shown that:

$$\Delta p_a(t_p) = \Delta p_{c,a}(t_p) + \Delta p_m = \frac{K_r^{p-1} S^{1-p}}{1 - e^{-S t_p}} [\Gamma(p) - \Gamma(p, S t_p)] + \mu J R_m \tag{10}$$

and

$$m_{c,a}(t_p) = Jc_b \left( \frac{1}{S} - \frac{t_p}{e^{St_p} - 1} \right) \tag{11}$$

where $\Delta p_a$ is the (total) TMP at process time $t_p$. $\Gamma(x, y)$ is the extended Euler gamma function defined by:

$$\Gamma(x, y) = \int_y^\infty \lambda^{x-1} e^{-\lambda} d\lambda \tag{12}$$

and

$$p = \frac{2-n}{1-n} \tag{13}$$

As mentioned previously, $n$ generally lies between 0.2 and 0.8 and thus $p$ is expected to be a positive quantity. The first term on the right-hand-side of Eq. (10), which increases with process time, is the contribution of cake buildup to the TMP while the second term, which remains constant, is that contributed by the membrane. It is to be noted from Eq. (11) that the transient cake mass $m_{c,a}$ is independent of the compressibility parameter $\alpha_0$ and viscosity $\mu$ of the permeate, and is only governed by the feed concentration $c_b$, permeate flux $J$, and surface-renewal rate $S$. Equation (10) contains the possibility of an inflection point occurring in the theoretical TMP profile, i.e., at a certain value of $t_p$, $d^2\Delta p_a / dt_p^2 = 0$.

We now examine the behavior of Eqs. (10) and (11) as $St_p \to 0$, i.e., as $S \to 0$ (low flow instability) or as $t_p \to 0$ (near the start of filtration). If one takes the limit of these expressions as $St_p \to 0$ (using L'Hôpital's rule), they give $\Delta p_a(St_p \to 0) = \mu J R_m$ and $m_{c,a}(St_p \to 0) = 0$, which are in accord with physical intuition. However, this method does not yield the time-dependent behavior of these quantities near $St_p = 0$, which can be deduced by means of the following procedure:

Differentiating Eq. (10) with respect to $t_p$ gives:

$$\frac{d\Delta p_a}{dt_p} = K_r^{p-1} S^{2-p} e^{-St_p} \left[ \frac{(St_p)^{p-1}(1 - e^{-St_p}) + \Gamma(p, St_p) - \Gamma(p)}{(1 - e^{-St_p})^2} \right] \quad (14)$$

from which it follows that as $St_p \to 0$:

$$\frac{d\Delta p_a}{dt_p}(St_p \to 0) = K_r^{p-1} t_p^{p-2} \quad (15)$$

Upon integrating Eq. (15), the following equation is obtained:

$$\Delta p_a(St_p \to 0) = \frac{K_r^{p-1}}{p-1} t_p^{p-1} + \Delta p_m \quad (16)$$

For an incompressible cake, $n = 0$ and thus $p = 2$ from Eq. (13). As $n$ varies from 0 to 0.8 (i.e., as cake compressibility increases), $p$ increases from 2 to 6.

***Variation of TMP with process time:*** At a fixed value of $K_r$, Eq. (15) indicates that the rate of change of the TMP with process or filtration time will be proportional to $t_p^{p-2}$ for $t_p \to 0$ or $S \to 0$. Also, according to Eq. (16), during the initial moments of filtration or for low levels of flow instability, the TMP will vary with the process time raised to a power of $p - 1$. Thus for $p = 2$ (incompressible cake), the TMP will increase linearly with process time as $St_p \to 0$, whereas for $p = 6$ (a highly compressible cake) it will increase as the 5th power of process time near the beginning of filtration or for low levels of flow instability, thus exhibiting very sharp concavity with respect to the process time (i.e., horizontal) axis.

***Variation of TMP with permeate flux:*** It can be observed from Eq. (2) that $K_r$ is proportional to the square of the permeate flux $J$. The rate of TMP increase with process time is proportional to $J^{2(p-1)}$ for $t_p \to 0$ or $S \to 0$, as can be inferred from Eq. (15). Thus, as $p$ changes from 2 to 6, (i.e., as the cake becomes more compressible), this rate will change proportionately as $J^2$ to $J^{10}$ for $St_p \to 0$, drastically increasing the concavity of the TMP versus process time profile.

The previous discussion indicates the extreme sensitivity of the shape of the TMP profile on the cake compressibility parameter $n$ and permeate flux $J$ during the initial moments of filtration or for low levels of flow instability. Such concave-type experimental TMP curves and the influence of permeate flux on the shape of the TMP profile can be observed in the data reported by Ho and Zydney[10] and Kovalsky et al.[11] for constant flux microfiltration; these data will be discussed later.

Differentiating Eq. (11) with respect to $t_p$ yields:

$$\frac{dm_{c,a}}{dt_p} = Jc_b \left[\frac{e^{St_p}(St_p - 1) + 1}{(e^{St_p} - 1)^2}\right] \qquad (17)$$

from which it follows that as $St_p \to 0$:

$$\frac{dm_{c,a}}{dt_p}(St_p \to 0) = Jc_b \qquad (18)$$

i.e.,

$$m_{c,a}(St_p \to 0) = Jc_b t_p \qquad (19)$$

Thus, according to Eq. (19), the mass of cake on the membrane surface will increase linearly with process time as $t_p \to 0$ or $S \to 0$.

It can be observed from Eqs. (16) and (19) that during the early moments of filtration or for low levels of flow instability, the TMP and mass of cake accumulated on the membrane surface are independent of the surface-renewal rate $S$, which does not appear in these equations. The absence of $S$ is a consequence of the Danckwerts age-distribution [Eq. (9)] that was used in our analysis; this distribution approaches a uniform distribution as $St_p \to 0$.

As $t_p \to \infty$, Eqs. (10) and (11) reduce to:

$$\Delta p_a(t_p \to \infty) = \Delta p_{lim} = \left(\frac{K_r}{S}\right)^{p-1} \Gamma(p) + \mu J R_m \qquad (20)$$

and

$$m_{c,a}(t_p \to \infty) = m_{c,lim} = \frac{Jc_b}{S} \qquad (21)$$

where $\Delta p_{lim}$ is the limiting or steady-state TMP and $m_{c,lim}$ is the mass of cake accumulated per unit area of the membrane surface when steady state is attained. According to Eqs. (20) and (21), both the limiting TMP and the accumulated mass of cake decrease as the level of flow instability, expressed by the magnitude of $S$, increases.

As $St_p \to \infty$, it can be shown from Eqs. (14) and (17) that:

$$\frac{d\Delta p_a}{dt_p}(St_p \to \infty) = 0 \qquad (22)$$

and

$$\frac{d\Delta m_{c,a}}{dt_p}(St_p \to \infty) = 0 \tag{23}$$

Thus as $St_p \to \infty$, the TMP and cake mass profiles will level out. This flattening of the profiles will occur earlier in the process for large values of $S$, i.e., there will be a quicker approach to steady state.

In principle, the four parameters of the complete model ($R_m$, $\alpha_0$, $n$, and $S$) can be estimated by the following procedure. The membrane resistance $R_m$ can be calculated from the equation:

$$R_m = \frac{1}{\mu P} \tag{24}$$

where $P$ is the pure-water permeance of the membrane, or it can be estimated from the experimental TMP value at the beginning of filtration. These two values of $R_m$ should be very close if the (initial) pore blocking of the membrane is negligible. For a finite level of flow instability, plotting Eq. (16) on logarithmic coordinates using initial experimental TMP versus process time data will allow the parameter $p$, and thus $n$ to be estimated [see Eq. (13)]. Different values of $S$ are now guessed and $\ln[\Delta p_a(t_p) - \mu J R_m]$ is plotted against $\ln[\{\Gamma(p) - \Gamma(p, St_p)\}/(1 - e^{-St_p})]$ using experimental TMP versus process time data [see Eq. (10)]. The value of $S$ which yields a magnitude of 1 for the average slope of the plot is the correct value of $S$. $K_r$ can then be calculated from the intercept of this plot [see Eq. (10)] after which $\alpha_0$ can be determined from Eq. (2). If the level of flow instability is low, Eq. (16) will hold for all values of the process time $t_p$. In this case, $K_r$ and $p$ (and thus $n$) can be determined from the intercept and slope of the experimental TMP data plotted on logarithmic coordinates [see Eq. (16)] after which $\alpha_0$ can be calculated from Eq. (2).

**Subsidiary Model: Incompressible Cake (*n* = 0)**

For *n* = 0 (i.e., *p* = 2), Eq. (10) reduces to:

$$\Delta p_a(t_p) = \Delta p_{c,a}(t_p) + \Delta p_m = K_r \left( \frac{1}{S} - \frac{t_p}{e^{St_p} - 1} \right) + \mu J R_m \qquad (25)$$

In the derivation of Eq. (25), the following relations have been used:

$$\Gamma(2) = 1 \qquad (26)$$

and

$$\Gamma(2, St_p) = (1 + St_p) e^{-St_p} \qquad (27)$$

Equation (27) is a special case of the formula:[15]

$$\Gamma(m, y) = \Gamma(m) \left[ 1 + y + \frac{y^2}{2!} + \cdots \frac{y^{m-1}}{(m-1)!} \right] e^{-y} \qquad (28)$$

where *m* is a positive integer.

Substituting *n* = 0 (i.e., *p* = 2) into Eqs. (16) and (20) yield:

$$\Delta p_a(St_p \to 0) = K_r t_p + \Delta p_m \qquad (29)$$

and

$$\Delta p_a(t_p \to \infty) = \Delta p_{lim} = \frac{K_r}{S} + \mu J R_m \qquad (30)$$

According to Eq. (29), $K_r$ is the slope of the TMP versus process time curve during the initial moments of filtration or for small levels of flow instability. The greater the values of $\mu$, $c_b$, $J$, and $\alpha_0$, the greater is the slope. Since $K_r$ is proportional to $J^2$ [Eq. (2)], the TMP will increase as the square of the permeate flux near $t_p = 0$ or as $S \rightarrow 0$. For example, an increase in $J$ by a factor of two will increase the slope by a factor of four. Equation (29) also shows that the TMP will increase linearly with process time during the early stages of filtration or for low levels of flow instability (as mentioned previously).

The growth in the mass of cake with process time is given by Eq. (11) while it's limiting behavior as $St_p \rightarrow 0$ or as $t_p \rightarrow \infty$ is given by Eq. (19) or (21).

The three parameters of the subsidiary model ($R_m$, $\alpha_0$ and $S$) can be determined by the following procedure. The membrane resistance $R_m$ can be calculated as for the complete model. For an assumed value of the surface-renewal rate $S$, $K_r$ can be estimated using Eq. (30) and the experimental value of $\Delta p_{lim}$, These values of $S$ and $K_r$ are then substituted into Eq. (25) and its fit to the experimental, transient TMP data is examined. This procedure is repeated for different (assumed) values of $S$ until the root-mean-square (RMS) deviation between the theoretical and experimental TMP is a minimum. At the end of this process, the 'best' value of $S$ will have been found after which $\alpha_0$ can be calculated from Eq. (2). If a value of $\Delta p_{lim}$ is not available (e.g., the experiment did not either reach steady state or was terminated before steady state was attained), $K_r$ may be estimated by fitting Eq. (29) to initial, experimental TMP versus process time data after which optimum values of $S$ and $\alpha_0$ can be determined as indicated earlier. However, this (extrapolation) method of extracting $K_r$ from initial TMP data may result in discrepancy between theory and experiment at large $t_p$.

The shape of the experimental TMP curve with process time will determine whether the complete model or the subsidiary model should be used. For convex-shaped TMP profiles which approach a plateau with the progress of process time, the subsidiary model should be adequate. If the TMP profile is initially concave and then becomes convex with the progress of process time, and subsequently levels out, the complete model is applicable, which should also be used if the TMP profile is concave for all values of the process time.

**RESULTS AND DISCUSSION**

As indicated earlier, Miller *et al.*[12] evaluated the use of unmodified and surface-modified membranes for constant flux CFMF. The reader is referred to their work for a detailed description of the materials used and experimental conditions and procedures, a very brief overview of which is provided in the following two paragraphs. A schematic of their constant flux CFMF system is also available in their paper.

The base (i.e., unmodified) ultrafiltration (UF) membrane material was (hydrophobic) polysulfone and came in molecular weight cutoffs of 10 and 20 kDa — these membranes were designated as PS-10 and PS-20, respectively. Two additional hydrophilic, surface-modified membranes were produced from PS-20, which were named PDA-modified and PDA75-modified. Polyethylene glycol (PEG) was grafted onto the surface of some of the PDA-modified sheets; such membranes were referred to as PDA-g-PEG- modified. The chief goal of such surface modification was to produce a hydrophilic surface on the base hydrophobic membrane for attracting water molecules that would act as a buffer between hydrophobic foulants and the membrane surface, thereby restricting their adsorption on it and also within the membrane pores. The steric hindrance offered by long PEG chains that extend from the membrane surface are also believed to further lessen the interaction between the surface and potential foulants. The pure-water permeance $P$ (measured by dead-end filtration) and estimated pore radius of the unmodified, PDA-modified, and PDA-g-PEG-modified PS-20 membranes are shown in Table 1, which also reports the $P$-value of the unmodified PS-10 membrane. It can be seen from this table that surface modification decreased the pure-water permeance of the PDA-modified and PDA-g-PEG-modified membranes by 22 and 37%, respectively, and also reduced the effective pore radius compared to the unmodified (PS-20) membrane.

The feed solution consisted of a 1500 ppm (1.5 kg m$^{-3}$) emulsion of soybean oil in water that had an average oil droplet size of 1.4 µm, with nearly all droplets lying in the size range of 0.8–3.0 µm. Thus, the average droplet size was two orders of magnitude greater than the effective pore radius of the unmodified (PS-20) membrane. The feed temperature was 25°C for all fouling experiments and the feed axial velocity was 0.18 m s$^{-1}$ (Reynolds number = 1000). The feed pressure was 2.1 barg (30 psig), which was maintained constant, and permeate and retentate were recycled back to the feed tanks. The membrane filtration area was 19.4 cm$^2$. The permeate flux was controlled at a constant rate by means of feedback control of a peristaltic pump installed on the permeate line. As the membrane fouled during an experimental run, the pressure on the permeate side of the membrane decreased, causing the TMP to increase. In cases of severe fouling, the pressure on the permeate side decreased to atmospheric pressure and the experiment was terminated. Membrane rejection was calculated by measuring the TOC (total organic carbon) content of the feed and permeate solutions. The membrane rejection values were quite high and were in the range of 96.5 – 99.1% with most them lying above 98%. Five constant flux levels were used in the experiments — 25, 40, 55, 70, 85, and 100 LMH (L m$^{-2}$ h$^{-1}$). A brief summary of the experimental parameters of Miller *et al.*[12] is provided in Table 2.

As indicated earlier, the feed suspension used by Miller *et al.*[12] contained soybean oil droplets whose average size (1.4 µm) was much larger than the effective pore radius of the unmodified (PS-20) membrane (4.2 nm). Also, as mentioned before, surface modification further reduced the pore size. It therefore can be conjectured that there was minimal or negligible pore blockage of the membranes in their experiments unless there was significant droplet breakage into much finer sizes or droplet deformation due to shearing forces and the applied TMP, and subsequent penetration into the pores of the membrane, especially at high permeate fluxes. That is, it can be postulated that the primary reason for the increase in TMP with process time observed in their experiments was cake formation on the membrane surface. It was therefore assumed that the membrane resistance $R_m$ could be calculated from Eq. (24) using values of the (average) membrane pure-water permeance $P$ reported by Miller *et al.*[12] (see Table 1) and using an estimated value of 8.98 × 10$^{-4}$ kg m$^{-1}$ s$^{-1}$ for the viscosity of water at 25°C (McCabe *et al.*[13]). This value was also used for the viscosity of the filtrate in the calculations. Values of $R_m$ for the three membranes are reported in Tab1e 3. Also, Miller *et al.*[12] state that "…all of the fouling experiments started, to good approximation, essentially instantaneously at a *V/A* value of zero." Here *V/A* is the ratio of the total or cumulative volume of permeate at process time $t_p$ to the filtration area and is a measure of process time, which can be obtained by dividing this ratio (reported by Miller *et al.*[12] in cm) by the permeate flux.

Figure 1 shows the fit of the subsidiary model [Eq. (25)] to the TMP data extracted from Fig. 2a of Miller *et al.*[12] for constant flux CFMF runs of soybean oil emulsion with the three membranes mentioned earlier at a permeate flux of 55 LMH ($J$ = 1.528 × 10$^{-5}$ m s$^{-1}$). It is observed that: (1) For all the membranes, the experimental TMP curve increases with process time rapidly at first and then at an extremely slow rate, and eventually approaches a plateau (i.e., steady state) as mentioned by Miller *et al.*[12], i.e., it is convex shaped. [In fitting the model to the data, it was assumed that the last experimental TMP value shown in the figure (at V/A ≈ 3.9 cm) was equal to $\Delta p_{lim}$, i.e., the limiting or steady-state TMP in the model.]. (2) The theoretical TMP curves begin at different levels of $\Delta p_m$ [pressure drop across the membrane; calculated from Eq. (5)] because of different values of the pure-water permeance *P* of the membranes (Table 1). (3) The unmodified (PS-20) membrane has the lowest TMP whereas the PDA-g-PEG-modified membrane has the highest, with that for the PDA-modified membrane being intermediate. (3) The unmodified (PS-20) and PDA-modified membranes have flatter TMP profiles compared to the PDA-g-PEG-modified membrane and show a faster approach to steady state. (4) There is a good fit of the model [Eq. (25)] to the experimental TMP data for the unmodified membrane while it is inferior in case of the modified membranes.

For the unmodified (PS-20) membrane, the average RMS deviation between the theoretical and experimental TMP is 2.9% with the surface-renewal rate $S$ being estimated at $4.2 \times 10^{-3}$ s$^{-1}$, which is comparable to values of $S$ reported by other researchers for cross-flow ultrafiltration and microfiltration.[2,6,8] The compressibility parameter $\alpha_0$ (i.e., specific cake resistance at a unit value of TMP), which measures the resistance offered by the accumulated material on the membrane surface to the flow of permeate and which depends on the packing density and nature of the cake, is estimated to be $9.07 \times 10^{13}$ m kg$^{-1}$, while the limiting or steady-state cake mass $m_{c,lim}$, calculated from Eq. (21), is found to be $5.46 \times 10^{-3}$ kg m$^{-2}$. If this value is multiplied by the membrane filtration area (19.4 cm$^2$), the total mass of accumulated solids on the membrane surface is calculated as 10.6 mg at steady-state. For the PDA-modified membrane the values are: $S = 8.5 \times 10^{-3}$ s$^{-1}$, $\alpha_0 = 2.53 \times 10^{14}$ m kg$^{-1}$ and $m_{c,lim} = 2.7 \times 10^{-3}$ kg m$^{-2}$. The mass of solids on the membrane surface at steady state = 5.2 mg while the RMS error of fit between theory and experiment = 5.8%, which is higher than the value of 2.9% obtained for the unmodified membrane. For the PDA-g-PEG-modified membrane, $S = 5.7 \times 10^{-3}$ s$^{-1}$, $\alpha_0 = 4.23 \times 10^{14}$ m kg$^{-1}$ and $m_{c,lim} = 4 \times 10^{-3}$ kg m$^{-2}$. The limiting mass of solids on the membrane surface = 7.8 mg and RMS error = 6.8%. For the benefit of the reader, values of the above model parameters, along with RMS deviations between theory and experiment, are gathered together in Table 3 for all three membranes.

Miller et al.[12] attributed the higher TMP of the modified membranes (compared to the unmodified membrane) to increased mass-transfer resistance due to the surface treatment of these membranes. However, in our opinion, surface treatment can only explain the higher (initial) resistance of these membranes (see Table 3), which will be manifested in an increased Δ$p_m$ at the beginning of filtration (Fig. 1). The subsequent increase in the TMP with process time for all the membranes is attributed to cake formation in our model [see Eq. (25)] while the differences in the rate of TMP increase for the three membranes are due to differences in the value of $α_0$. Table 3 indicates that the values of $R_m$ for the PDA-modified and PDA-g-PEG-modified membranes are 29 and 58% higher while values of $α_0$ for these membranes are 179 and 367% greater, respectively, compared to the corresponding values for the unmodified (PS-20) membrane. This partially explains the high TMP of the PDA-g-PEG-modified membrane, the intermediate TMP of the PDA-modified membrane and the low TMP of the unmodified (PS-20) membrane. Figure 2 exhibits the theoretical, age-averaged cake-mass profile [calculated from Eq. (11)] for the three membranes. All the curves in this figure start from a value of zero and, with the progress of filtration, approach the corresponding steady-state values reported earlier. Although the axial velocity of the feed suspension was maintained at the same level (0.18 m s$^{-1}$) in the experimental runs of Miller *et al.*,[12] it is observed from Table 3 that the values of the surface-renewal rate $S$, which depend upon the prevailing hydrodynamic conditions near the membrane surface, are different for the three membranes. These differences may be speculated as being due to differences in membrane surface roughness amongst the three membranes since the roughness will have an effect on the micro-scale hydrodynamics. However, no definite conclusion can be drawn because of the large variance of the permeance about its mean value for all three membranes as can be seen from Table 1. We attempted to fit the TMP data by changing the permeance within its variance for all three membranes and were able to fit them with values of $S$ that were more uniform, but with somewhat larger RMS errors. The greater the value of $S$, the flatter is the TMP profile and smaller is the cake accumulation on the membrane. As can be seen from Table 3 and Fig. 2, the trend of variation of $S$ among the membranes corresponds to the trend of variation of the cake mass deposited on the membrane surface.

The results discussed so far concerned the performance of membranes which had different pure-water permeances. In order to compare membrane performance on an equivalent basis, Miller *et al*.[12] performed experimental runs with PS-10 and PDA75-modified (PS-20) membranes. These membranes had a pure-water permeance of 570 LMH bar$^{-1}$, which was the same as that of the PDA-g-PEG-modified (PS-20) membrane. Figure 3 shows the experimental TMP data (extracted from Fig. 6 of Miller *et al*.[12]) for these membranes at a permeate flux of 55 LMH with all other experimental conditions, as described earlier, remaining the same. It is also observed in this figure that there is a fairly good fit of the subsidiary model to the convex-type experimental TMP profile for the unmodified membrane but substantial errors result in case of the modified membranes as can be seen from the RMS deviations in Table 4, which also reports values of the model parameters. This can be seen more clearly in Fig. 3A, which shows part of the data of Fig. 3 in expanded scale. The subsidiary model [Eq. (25)] is unable to quite capture the initially convex and subsequent slow rise of the experimental TMP for the modified membranes. Also, in contrast to Fig. 1, it is the unmodified PS-10 membrane that has a much higher TMP compared to the surface-modified membranes, whose TMP curves are very close to one another and much flatter. The superior performance of the modified membranes can be attributed, as indicated by Miller *et al.*,[12] to the beneficial effects of hydrophilicity and steric hindrance that retard and counteract the accumulation of solids on the membrane surface, which are driven to it by the flow of liquid. Such effects are manifested in a lower value of $α_0$ for a surface-modified membrane. Thus, values of this parameter are $5.17 \times 10^{14}$, $4.1 \times 10^{14}$ and $4.34 \times 10^{14}$ m kg$^{-1}$ for the unmodified PS-10, PDA75-modified (PS-20) and PDA-g-PEG-modified (PS-20) membranes, respectively (Table 4). Figure 4 compares the theoretical cake-mass buildup on the membrane surface for the three membranes. In contrast to Fig. 2, the order of the cake-mass buildup curves in this figure follow the order of the corresponding theoretical TMP curves in Fig. 3, with the modified membranes accumulating significantly lesser amounts of cake compared to the unmodified membrane due to their higher surface-renewal rates (Table 4).

The average RMS error of the fit of the subsidiary model to the experimental TMP data of Miller *et al.*,[12] shown in Figs. 1 and 3, is 6.2%. The inferior fit of this model to the data in case of the modified membranes compared to that for the unmodified membrane can be postulated to arise from hydrophilicity, steric hindrance and other surface (e.g., electrochemical) effects that result from modifying the surface of the membrane, which are not explicitly accounted for in the surface-renewal model. Miller *et al.*[12] also reported TMP data at higher permeate fluxes of 70 and 85 LMH (see Figs. 2b and 2c in their paper) when cake-compressibility effects are expected to become important. At a flux of 70 LMH, the experimental TMP curves for the PDA-modified and PDA-g-PEG modified membranes still exhibited convex-type behavior. However, the TMP curve for the unmodified (PS-20) membrane was initially convex shaped until a certain value of the *V/A* ratio after which it became concave and rose rapidly, i.e., an inflection point can be clearly observed in the experimental TMP curve. At a flux of 85 LMH, the TMP curves for all three membranes exhibited this latter type of behavior with crisscrossing of the curves. According to Miller *et al.*,[12] this behavior occurs when the threshold flux of the membrane is crossed, which brings about the onset of intense fouling. We were unsuccessful in accounting for this complex TMP behavior with the surface-renewal model developed in this work.

Ho and Zydney[10] studied the microfiltration of BSA solutions in a 25-mm diameter stirred ultrafiltration cell using polycarbonate track-etched (PCTE) membranes with two different values of porosity (3 and 10%). The concentration of BSA in the feed solution was 2 kg m$^{-3}$. Assuming gentle stirring (i.e., $S \rightarrow 0$) and an experimental temperature (not reported in their work) of 25°C, we attempted to fit their experimental TMP versus process time data with Eq. (16) of the complete model. Figure 5 shows model comparisons with their experimental TMP profiles for a PCTE membrane (porosity of 10%) for three different permeate fluxes. The experimental TMP builds up slowly at first after which there is a rapid increase due to cake buildup on the membrane surface. The model is able to capture the concave shape of the experimental TMP profile, which shows a very marked sensitivity to the level of the imposed permeate flux as anticipated earlier in the theoretical section. Values of the three model parameters ($R_m$, $\alpha_0$ and $n$) are reported in Table 5 along with the RMS error, whose average value is 11.5%. The value of $R_m$ varies somewhat between the membranes with an average value of 3.4 × $10^{10}$ m$^{-1}$. Since a separate clean membrane was used by Ho and Zydney[10] for each experiment, the variability in $R_m$ can be attributed to variability in the surface characteristics of the individual PCTE membrane sheets. The individual values of $\alpha_0$ are of the same order of magnitude with an average value of 1.91 × $10^9$ m kg$^{-1}$ whereas the values of $n$ are quite consistent with an average value of 0.66. These are significantly different from the values of $\alpha_0$ = 1.7 ± 0.02 × $10^{12}$ m kg$^{-1}$ and $n$ = 0.78 ± 0.01 obtained by Ho and Zydney[10] by fitting their 5-parameter numerical model to these same data. These 5 parameters are: membrane resistance (whose values they did not report), resistance of a single protein aggregate, pore-blockage parameter, and the two compressibility parameters ($\alpha_0$ and $n$). Through independent measurements (see below), they estimated values of $\alpha_0$ and $n$ to be 3 × $10^{12}$ m kg$^{-1}$ and 0.82, respectively, which compare well with their values given previously.

The following explanation is offered to account for the discrepancy between the values of the compressibility parameters ($\alpha_0$ and $n$) reported by Ho and Zydney[10] and those obtained in this paper by fitting the surface-renewal model to their TMP data. Using the resistance-in-series model of constant pressure microfiltration, Ho and Zydney[16] calculated the specific cake resistance $\alpha$ from the following two equations:

$$J_{lim} = \frac{\Delta p}{\mu R_{tot}} \qquad (31)$$

with

$$R_{tot} = R_m + \alpha m_p \tag{32}$$

Here $m_p$ is the (steady-state) mass of the protein (i.e., cake) layer deposit per unit area of the membrane surface whereas $R_{tot}$ is the total resistance (membrane plus cake). By measuring the steady-state saline flux through a heavily fouled membrane and the difference in weights of the clean and fouled membrane (for estimating $m_p$), Ho and Zydney[16] calculated values of $\alpha$ at different TMP values from Eqs. (31) and (32) after which $\alpha_0$ and $n$ were determined by plotting Eq. (3) on logarithmic coordinates. It is to be noted that, according to Eq. (31), the steady-state permeate flux is directly proportional to $\Delta p / R_{tot}$. However, according to the constant pressure, surface-renewal model of microfiltration (Hasan *et al.*[8]):

$$J_{lim} = \sqrt{\frac{\pi S \Delta p}{2\mu c_b \alpha}} = \sqrt{\frac{\pi S \Delta p^{1-n}}{2\mu c_b \alpha_0}} = \sqrt{\frac{\pi S}{2\mu c_b \alpha_0{}^n}} \sqrt{\left(\frac{\Delta p}{\alpha_0}\right)^{1-n}} \tag{33}$$

Thus, according to Eq. (33), the limiting permeate flux is proportional to $\sqrt{\Delta p / \alpha}$ [assuming all other parameters in Eq. (33) to remain constant]. Values of $\alpha_0$ can now be guessed and the experimental values of $\ln J_{lim}^2$ plotted against $\ln(\Delta p / \alpha_0)$. That value of $\alpha_0$ which yields a straight line on this plot is the correct value of $\alpha_0$, with the slope of the line being equal to $1 - n$. As is evident, the theoretical framework of the surface-renewal model is radically different from that of the resistance-in-series model, which explains the discrepancy between the values of $\alpha_0$ and $n$ obtained in this work and those reported by Ho and Zydney.[10] Thus, these values are not absolute quantities but depend upon the theoretical framework used to analyze the experimental data.

The theoretical cake-mass buildup as a function of process or filtration time [calculated from Eq. (19)] corresponding to the three permeate fluxes in Fig. 5 is shown in Fig. 6. The buildup occurs in a linear fashion and the higher the permeate flux, the greater is its magnitude at a specified value of the process time.

Figure 7 compares the complete model with experimental TMP profiles of PCTE and PCTE-L (porosity of 3%) membranes at an imposed permeate flux of $1.3 \times 10^{-4}$ m/s. Table 6 reports the values of the model parameters and RMS errors (average RMS error = 10.9%). The values of $\alpha_0$ and $n$ for the PCTE membrane are $4.2 \times 10^8$ m kg$^{-1}$ and 0.76, respectively, which are appreciably different from the (average) values of $1.91 \times 10^9$ m kg$^{-1}$ and 0.66 estimated earlier for this type of membrane. In order to understand the reason for this discrepancy, all the experimental TMP data of Ho and Zydney[10] are plotted together in Fig. 8. The experimental TMP profiles for the permeate fluxes of $1.1 \times 10^{-4}$ and $1.3 \times 10^{-4}$ m/s virtually coincide for the first 25 min of filtration after which they diverge from one another. After this point in time, the TMP profile for the flux of $1.3 \times 10^{-4}$ m/s rises in a more or less parallel fashion to that for the flux of $1.5 \times 10^{-4}$ m/s. It can be conjectured that this behavior is due to variability of the individual PCTE membrane sheets and/or some inconsistency in the experimental procedure.

Figure 9 shows the linear growth of the mass of cake with filtration time on the surface of the membrane (PCTE and PCTE-L) for the permeate flux of $1.3 \times 10^{-4}$ m/s.

The theoretical TMP curves of Ho and Zydney[10] shown in Figs. 1 and 2 of their paper exhibit convex regions superimposed on the overall concave shape of the TMP profiles at intermediate to large values of the process time, which cannot be discerned in their experimental TMP data.

Finally, Fig. 10 compares the complete model [Eq. (16)] against the experimental TMP data of Kovalsky et al.[11] for filtration of a 10 kg m$^{-3}$ yeast suspension in a stirred cell (filtration area = 19.6 cm$^2$; stirring speed = 5 RPM) for three different values of the imposed permeate flux. A temperature of 25°C was assumed in the theoretical calculations. Once again it is seen that the model, whose parameters are reported in Table 7, is able to capture the concave shape of the experimental TMP profile fairly (average RMS error = 9.2%; see Table 7). The discrepancy between model and experiment near the end of an experimental run can be attributed to creep and consolidation effects, which were considered in the numerical model of Kovalsky et al.[11].

The linear growth in the mass of cake with filtration time is shown in Fig. 11.

**CONCLUSIONS**

This paper presented a mathematical model of constant flux CFMF by combining classical cake-filtration theory with the surface-renewal concept. The model can predict the TMP development and cake buildup on the membrane surface with filtration time. The basic parameters of the model are the membrane resistance, specific cake resistance, and rate of surface renewal. There are two versions of the surface-renewal model: the complete model, which accounts for cake compressibility, and a subsidiary model which can be derived from the complete model when the cake is incompressible. The subsidiary model was correlated against some of the experimental TMP data recently reported by Miller *et al*.[12] for constant flux CFMF of a soybean-oil emulsion in a cross-flow filtration cell having unmodified and surface-modified, fouling-resistant membranes. Although the average RMS error of the fit was 6.2%, the quality of the fit was much better for the unmodified membrane. The complete model was fitted to the constant flux, stirred-cell, BSA microfiltration TMP data of Ho and Zydney[10] and also to the TMP data of Kovalsky *et al*.[11] for yeast filtration in a stirred cell. The average RMS errors of the fit were 11.2 and 9.2%, respectively.

The essence of the surface-renewal model is its ability to explicitly account for flow instabilities generated at the membrane surface (due to membrane roughness, presence of spacers, etc.) through the hydrodynamic parameter $S$, which is in contrast to the other models of membrane filtration (e.g., the film, boundary-layer or resistance models). As demonstrated in this work, the model has the ability to correlate convex- and concave-shaped experimental TMP profiles but may not be suitable for representing more complex TMP behavior. Unlike the CFMF models of Ho and Zydney[10] and Kovalsky *et al*.,[11] the surface-renewal model provides explicit, analytical expressions for the TMP and cake-mass buildup on the membrane surface as a function of filtration time. For future work it is suggested that the model be rigorously tested for its ability to predict the influence of feed concentration and axial liquid velocity on the TMP and also be extended to account for the effects of hydrophilicity, steric hindrance, etc. that result from surface modification of membranes. Incorporating the phenomena of pore blocking and cake consolidation into the model would make it more widely applicable.

**ACKNOWLEDGEMENT**

S.G.C. thanks Mr. Susumu Ikuta and Dr. Noshir Mistry for thought provoking discussions on TMP behavior and membrane fouling in constant flux, cross-flow microfiltration.

**NOMENCLATURE**

| | | |
|---|---|---|
| $A$ | filtration area of membrane | cm² or m² |
| $c_b$ | mass of solids deposited in the filter per unit volume of filtrate (approximately equal to the concentration of solids in the feed or bulk liquid) | kg m$^{-3}$ |
| $f(t, t_p)$ | age-distribution function of liquid elements at the membrane wall | s$^{-1}$ |
| $J$ | constant permeate flux | m s$^{-1}$ |
| $K_r$ | defined by Eq. (2) | kg m$^{-1}$ s$^{-3}$ |
| $m$ | positive integer (1, 2, 3, …) | |
| $m_c$ | mass of cake in a liquid element per unit area of the membrane surface at time $t$ | kg m$^{-2}$ |
| $m_{c,a}$ | age-averaged mass of cake per unit area of the membrane surface at process time $t_p$ | kg m$^{-2}$ |
| $m_{c,lim}$ | limiting or steady-state mass of cake per unit area of the membrane surface | kg m$^{-2}$ |
| $m_p$ | steady-state mass of protein layer deposit per unit area of the membrane surface | kg m$^{-2}$ |
| $n$ | compressibility coefficient of the cake | |
| $p$ | defined by Eq. (13) | |
| $P$ | Pure-water permeance of the membrane | L m$^{-2}$ h$^{-1}$ bar$^{-1}$ or m$^2$ s kg$^{-1}$ |
| $R_m$ | hydraulic resistance of the membrane | m$^{-1}$ |
| $R_{tot}$ | total resistance (membrane plus cake) defined by Eq. (32) | m$^{-1}$ |
| $S$ | rate of renewal of liquid elements at the membrane surface | s$^{-1}$ |
| $t$ | residence time of a liquid element at the membrane surface | s |
| $t_p$ | filtration or process time | s |
| $V$ | cumulative or total volume of permeate at process time $t_p$ | m$^3$ |
| $x$ | parameter of $\Gamma(x, y)$ | |
| $y$ | parameter of $\Gamma(x, y)$ | |

**Greek Symbols**

| | | |
|---|---|---|
| $\alpha$ | specific cake resistance | m kg$^{-1}$ |
| $\alpha_0$ | compressibility parameter of the cake (or specific cake resistance per unit transmembrane pressure drop) | m kg$^{-1}$ |
| $\Gamma(x, y)$ | extended Euler gamma function; defined by Eq. (12) | |
| $\Delta p$ | (total) transmembrane pressure drop in a surface element at time $t$ [Eq. (4)] | Pa |
| $\Delta p_a$ | age-averaged (total) transmembrane pressure drop at process time $t_p$ | Pa |
| $\Delta p_c$ | pressure drop across the cake in a surface element at time $t$ | Pa |
| $\Delta p_{c,a}$ | age-averaged pressure drop across the cake at process time $t_p$ [Eq. (7)] | Pa |
| $\Delta p_{lim}$ | limiting or steady-state (total) transmembrane pressure drop | Pa |
| $\Delta p_m$ | pressure drop across the membrane | Pa |
| $\lambda$ | variable of integration in Eq. (12) | |
| $\mu$ | viscosity of the permeate | kg m$^{-1}$ s$^{-1}$ |

**REFERENCES**

1. Koltuniewicz, A. Predicting permeate flux in ultrafiltration on the basis of surface renewal concept, *J. Membrane Sci.* **1992**, *68*, 107-118.

2. Koltuniewicz, A.; Noworyta, A. Dynamic properties of ultrafiltration systems in light of the surface renewal theory, *Ind. Eng. Chem. Res.* **1994**, *33,* 1771-1779.

3. Koltuniewicz, A.; Noworyta, A. Method of yield evaluation for pressure-driven membrane processes, *Chem. Eng. J.* **1995**, *58*, 175-182.

4. Constenla, D. T.; Lozano, J. E. Predicting stationary permeate flux in the ultrafiltration of apple juice, *Lebensm. Wiss. u. Technol.* **1996**, *29*, 587-592.

5. Arnot, T. C.; Field, R. W.; Koltuniewicz, A. B. Cross-flow and dead-end microfiltration of oily-water emulsions. Part II: Mechanisms and modeling of flux decline, *J. Membrane Sci.* **2000**, *169*, 1-15.

6. Chatterjee, S. G. On the use of the surface-renewal concept to describe cross-flow ultrafiltration, *Indian Chemical Engineer* **2010**, *52*, 179-193.

7. Sarkar, D.; Datta, D.; Sen, D.; Bhattacharjee, C. Simulation of continuous stirred rotating disk-membrane module: An approach based on surface renewal theory, *Chem. Eng. Sci.* **2011**, *66*, 2554-2567.

8. Hasan, A.; Peluso, C. R.; Hull, T. S.; Fieschko, J.; Chatterjee, S. G. A surface-renewal model of cross-flow microfiltration, *Brazilian Journal of Chemical Engineering* **2013**, *30*, 167-186.

9. Zhang, W.; Chatterjee, S. G. Influence of residence time distribution on a surface-renewal model of constant-pressure cross-flow microfiltration, *Brazilian Journal of Chemical Engineering* **2014**, accepted for publication.

10. Ho, C-C.; Zydney, A. L. Transmembrane pressure profiles during constant flux microfiltration of bovine serum albumin, *J. Membrane Sci.* **2002**, *209*, 363-377.

11. Kovalsky, P.; Bushell, G.; Waite, T. D. Prediction of transmembrane pressure build-up in constant flux microfiltration of compressible materials in the absence and presence of shear, *J. Membrane Sci.* **2009**, *344*, 204-210.

12. Miller, D. J.; Kasemset, S.; Wang, L.; Paul, D. R.; Freeman, B. D. Constant flux crossflow filtration evaluation of surface-modified fouling-resistant membranes, *J. Membrane Sci.* **2014**, *452*, 171-183.

13. McCabe, W. L.; Smith, J. C.; Harriott, P. Unit Operations of Chemical Engineering; McGraw-Hill: New York, **1993**; Fifth Ed.

14. Danckwerts, P. V. Significance of liquid-film coefficients in gas absorption. *Ind. Eng. Chem. (Eng. and Process Dev.)* 1951, *43*, 1460-1467.

15. Abramowitz, M.; Stegun, I. A., Eds.; In Handbook of Mathematical Functions — With Formulas, Graphs, and Mathematical Tables; Dover Publications: New York, **1965**; Chapter 6, p. 262.

16. Ho, C-C.; Zydney, A. L. A combined pore blockage and cake filtration model for protein fouling during microfiltration, *J. Colloid and Interface Sci.* **2000**, *232*, 389-399.

TABLE 1 Membrane Characteristics of Miller *et al.*[12]

| Membrane | Pure water permeance (LMH bar$^{-1}$) | Effective pore radius (nm) |
|---|---|---|
| Unmodified (PS-20) | 900 ± 200 | 4.2 |
| PDA-modified (PS-20) | 700 ± 100 | 3.7 |
| PDA-g-PEG-modified (PS-20) | 570 ± 70 | 3.3 |
| PDA75-modified (PS-20) | 570 | – |
| Unmodified (PS-10) | 570 | – |

TABLE 2 Experimental Parameters of Miller et al.[12]

| Parameter | Description or value |
|---|---|
| Base membrane type | UF polysulfone with molecular weight cutoffs of 10 and 20 kDa |
| Type of feed suspension | Soybean-oil emulsion in water |
| Droplet average size in feed suspension | 1.4 µm |
| Feed concentration | 1.5 kg m$^{-3}$ |
| Feed pressure | 2.1 barg (30 psig) |
| Membrane filtration area | 19.4 cm$^2$ |
| Feed axial velocity | 0.18 m s$^{-1}$ (Reynolds number = 1000) |
| Experimental temperature | 25°C |
| Constant flux levels | 25, 40, 55, 70, 85, and 100 LMH (L m$^{-2}$ h$^{-1}$) |

TABLE 3 Parameter Values of the Subsidiary Model [Eq. (25)] and RMS Deviations between Theory and Experiment for the TMP Data shown in Fig. 2a of Miller *et al.*[12]

| Parameter | Unmodified (PS-20) | PDA-modified (PS-20) | PDA-g-PEG-modified (PS-20) |
|---|---|---|---|
| $R_m$ (m$^{-1}$) | $4.45 \times 10^{11}$ | $5.73 \times 10^{11}$ | $7.03 \times 10^{11}$ |
| $S$ (s$^{-1}$) | $4.2 \times 10^{-3}$ | $8.5 \times 10^{-3}$ | $5.7 \times 10^{-3}$ |
| $\alpha_0$ (m kg$^{-1}$) | $9.07 \times 10^{13}$ | $2.53 \times 10^{14}$ | $4.23 \times 10^{14}$ |
| $n$ | 0 | 0 | 0 |
| RMS deviation (%) | 2.9 | 5.8 | 6.8 |

TABLE 4 Parameter Values of the Subsidiary Model [Eq. (25)] and RMS Deviations between Theory and Experiment for the TMP Data shown in Fig. 6 of Miller *et al.*[12]

| Parameter | Unmodified (PS-10) | PDA75-modified (PS-20) | PDA-g-PEG-modified (PS-20) |
|---|---|---|---|
| $R_m$ (m$^{-1}$) | $7.03 \times 10^{11}$ | $7.03 \times 10^{11}$ | $7.03 \times 10^{11}$ |
| $S$ (s$^{-1}$) | $1.8 \times 10^{-3}$ | $4.8 \times 10^{-3}$ | $5.6 \times 10^{-3}$ |
| $\alpha_0$ (m kg$^{-1}$) | $5.17 \times 10^{14}$ | $4.1 \times 10^{14}$ | $4.34 \times 10^{14}$ |
| $n$ | 0 | 0 | 0 |
| RMS deviation (%) | 3.3 | 12 | 6.3 |

TABLE 5 Parameter Values of the Complete Model [Eq. (16)] and RMS Deviations between Theory and Experiment for the TMP Data shown in Fig. 1 of Ho and Zydney.[10]

| Parameter | $J = 0.7 \times 10^{-4}$ (m/s) | $J = 1.1 \times 10^{-4}$ (m/s) | $J = 1.5 \times 10^{-4}$ (m/s) |
|---|---|---|---|
| $R_m$ (m$^{-1}$) | $2.84 \times 10^{10}$ | $3.47 \times 10^{10}$ | $3.89 \times 10^{10}$ |
| $\alpha_0$ (m kg$^{-1}$) | $2.26 \times 10^{9}$ | $1.90 \times 10^{9}$ | $1.57 \times 10^{9}$ |
| $n$ | 0.67 | 0.65 | 0.65 |
| RMS deviation (%) | 16.8 | 9.6 | 8.1 |

TABLE 6 Parameter Values of the Complete Model [Eq. (16)] and RMS Deviations between Theory and Experiment for the TMP Data shown in Fig. 2 of Ho and Zydney.[10]

| Parameter | PCTE-L | PCTE |
|---|---|---|
| $R_m$ (m$^{-1}$) | $1.71 \times 10^{11}$ | $4.37 \times 10^{10}$ |
| $\alpha_0$ (m kg$^{-1}$) | $1.12 \times 10^{10}$ | $4.20 \times 10^{8}$ |
| $n$ | 0.57 | 0.76 |
| RMS deviation (%) | 7.6 | 14.1 |

TABLE 7 Parameter Values of the Complete Model [Eq. (16)] and RMS Deviations between Theory and Experiment for the TMP Data shown in Fig. 5 of Kovalsky et al.[11]  LMH = L/m²/h.

| Parameter | 30 LMH | 40 LMH | 50 LMH |
|---|---|---|---|
| $R_m$ (m$^{-1}$) | $1.63 \times 10^{12}$ | $1.17 \times 10^{12}$ | $1.02 \times 10^{12}$ |
| $\alpha_0$ (m kg$^{-1}$) | $1.18 \times 10^{9}$ | $9.10 \times 10^{9}$ | $5.92 \times 10^{9}$ |
| $n$ | 0.75 | 0.58 | 0.62 |
| RMS deviation (%) | 10.5 | 8.8 | 8.4 |

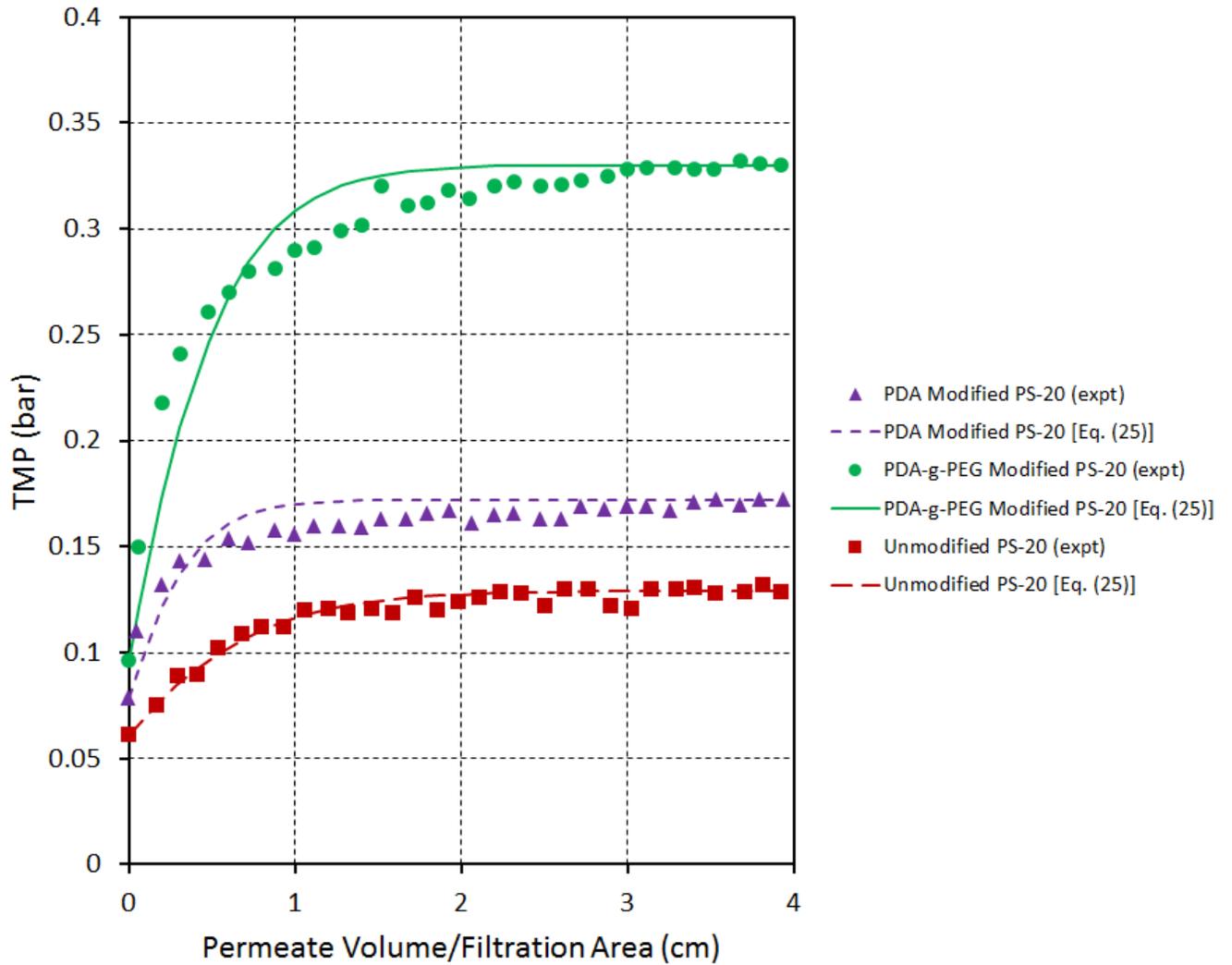

FIGURE 1 Comparison of the Subsidiary Model [Eq. (25)] and Experimental (Fig. 2a of Miller et al.[12]) TMP Profiles in the Microfiltration of a Soybean-Oil Suspension. Values of Experimental and Model Parameters are provided in Tables 1, 2 and 3. Permeate Flux = 55 LMH (L/m$^2$/h).

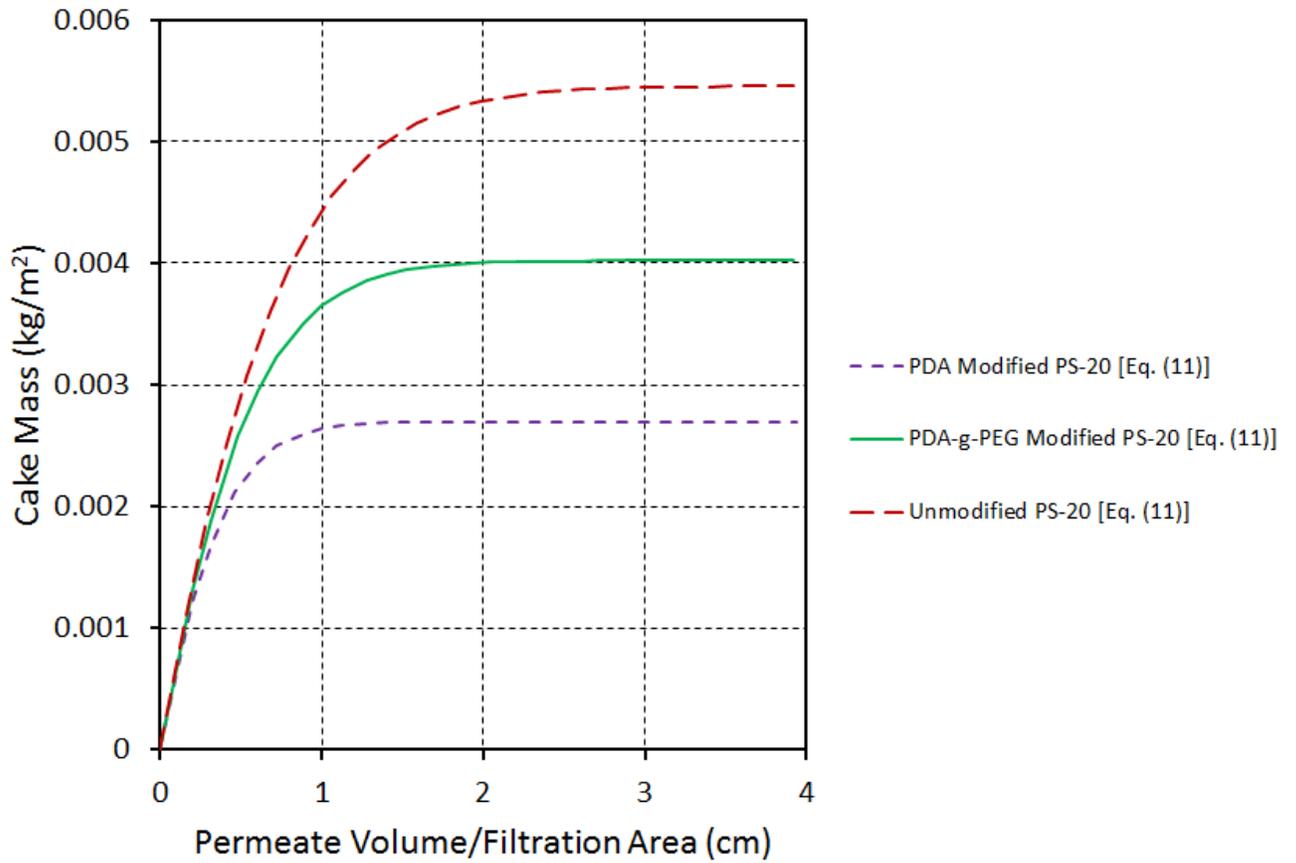

FIGURE 2 Predicted Cake Buildup [Eq. (11)] for the Experiments of Miller *et al.*[12] in the Microfiltration of a Soybean-Oil Suspension. Values of Experimental and Model parameters are provided in Tables 1, 2 and 3. Permeate Flux = 55 LMH (L/m$^2$/h).

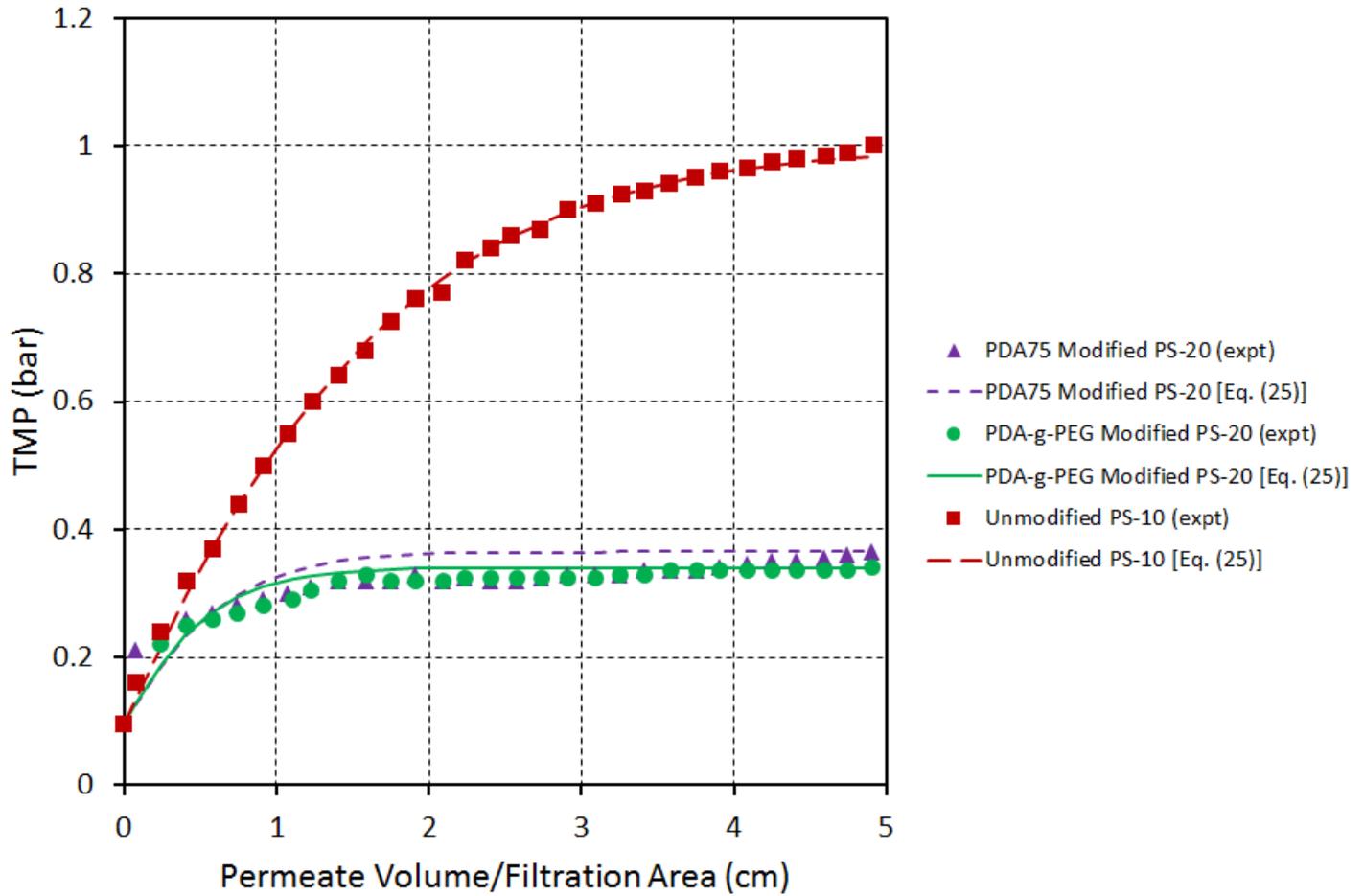

FIGURE 3 Comparison of the Subsidiary Model [Eq. (25)] and Experimental (Fig. 6 of Miller et al.[12]) TMP Profiles in the Microfiltration of a Soybean-Oil Suspension. Values of Experimental and Model Parameters are provided in Tables 1, 2 and 4. Permeate Flux = 55 LMH (L/m$^2$/h).

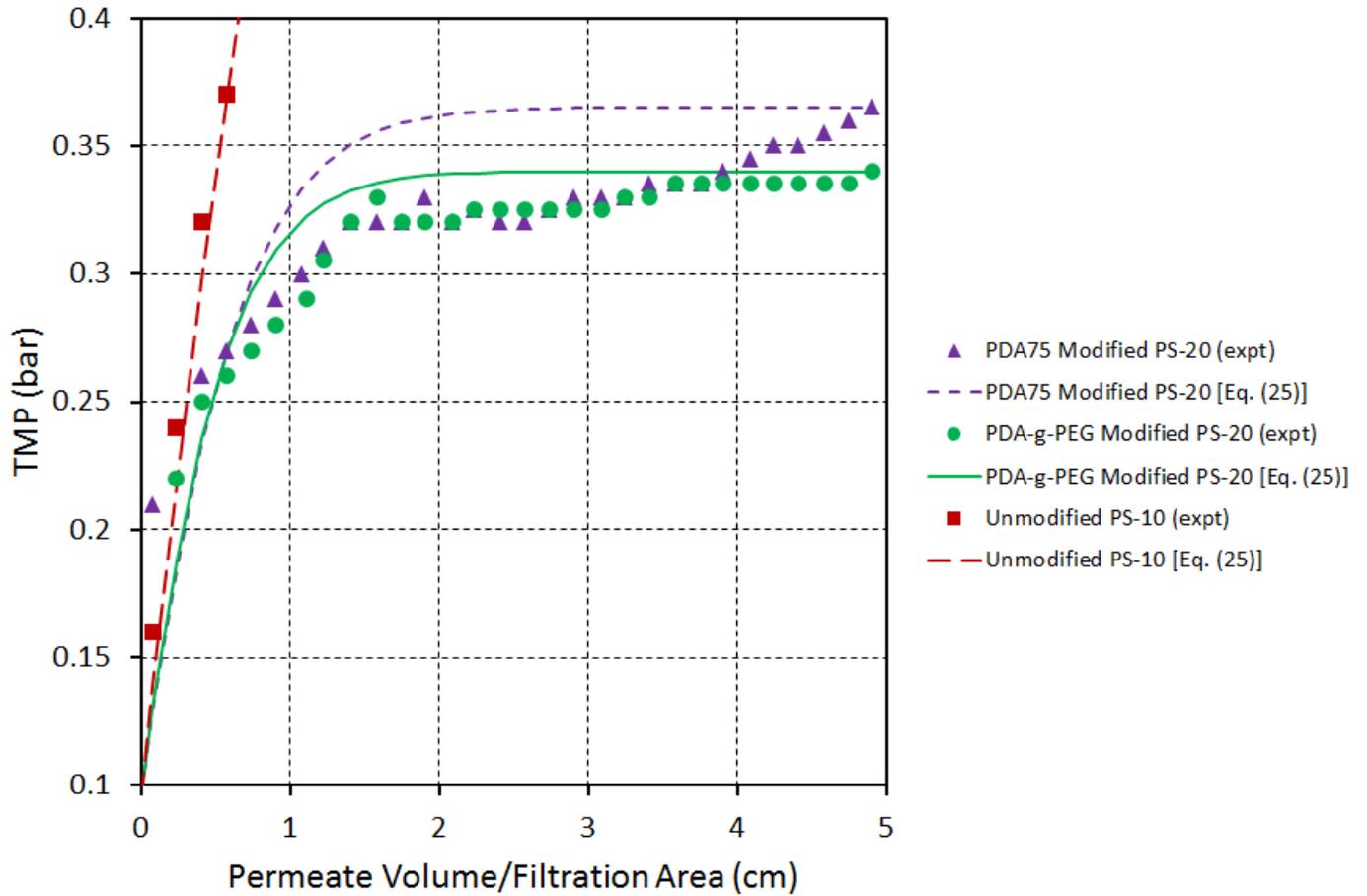

FIGURE 3A Comparison of the Subsidiary Model [Eq. (25)] and Experimental (Fig. 6 of Miller *et al.*[12]) TMP Profiles in the Microfiltration of a Soybean-Oil Suspension — Expanded Scale. Values of Experimental and Model Parameters are provided in Tables 1, 2 and 4. Permeate Flux = 55 LMH (L/m$^2$/h).

FIGURE 4 Predicted Cake Buildup [Eq. (11)] for the Experiments of Miller et al.[12] in the

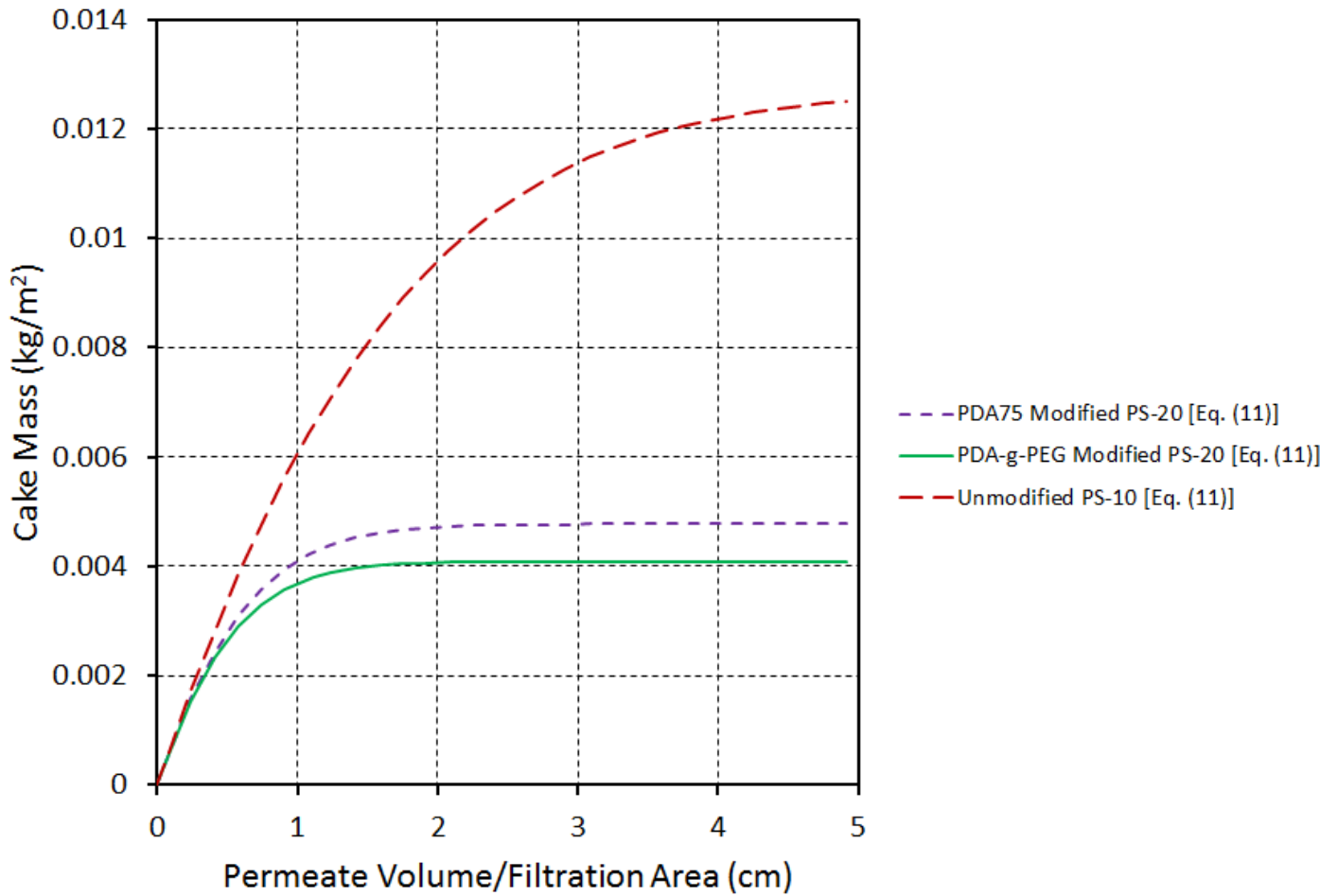

Microfiltration of a Soybean-Oil Suspension. Values of Experimental and Model Parameters are provided in Tables 1, 2 and 4. Permeate Flux = 55 LMH (L/m$^2$/h).

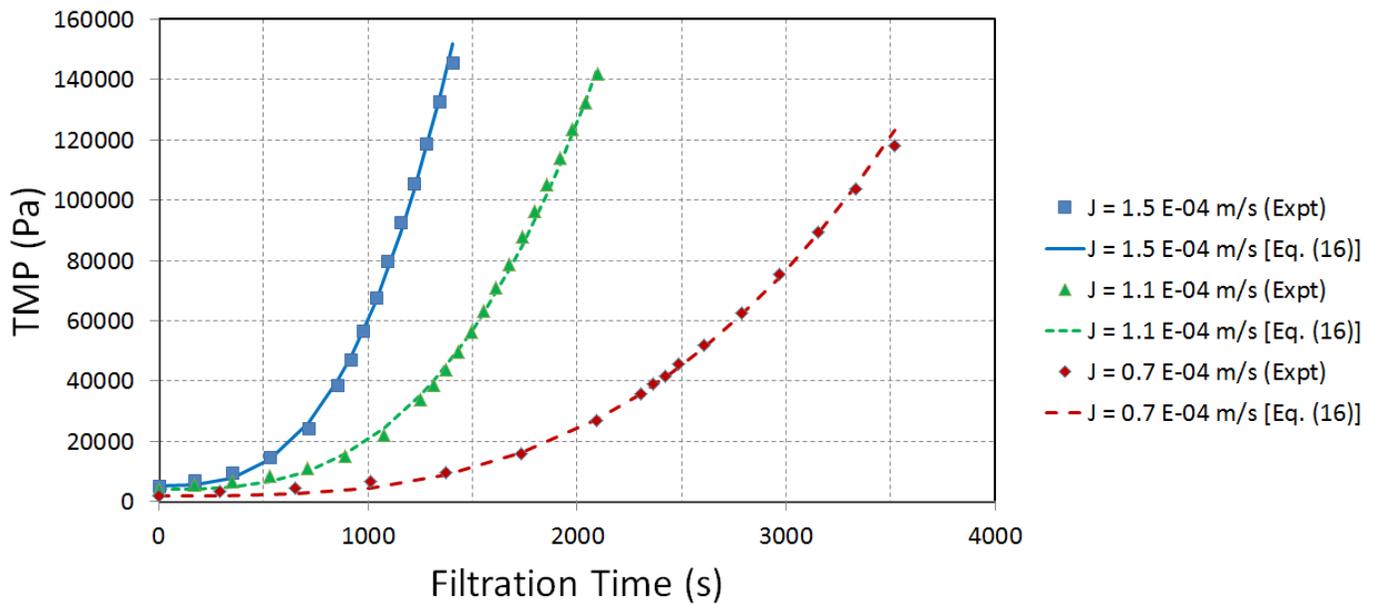

FIGURE 5 Comparison of the Complete Model [Eq. (16)] and Experimental (Fig. 1 of Ho and Zydney[10]) TMP Profiles in the Microfiltration of a BSA Solution with a PCTE Membrane. Values of Model Parameters are provided in Table 5. Feed Concentration = 2 kg/m$^3$.

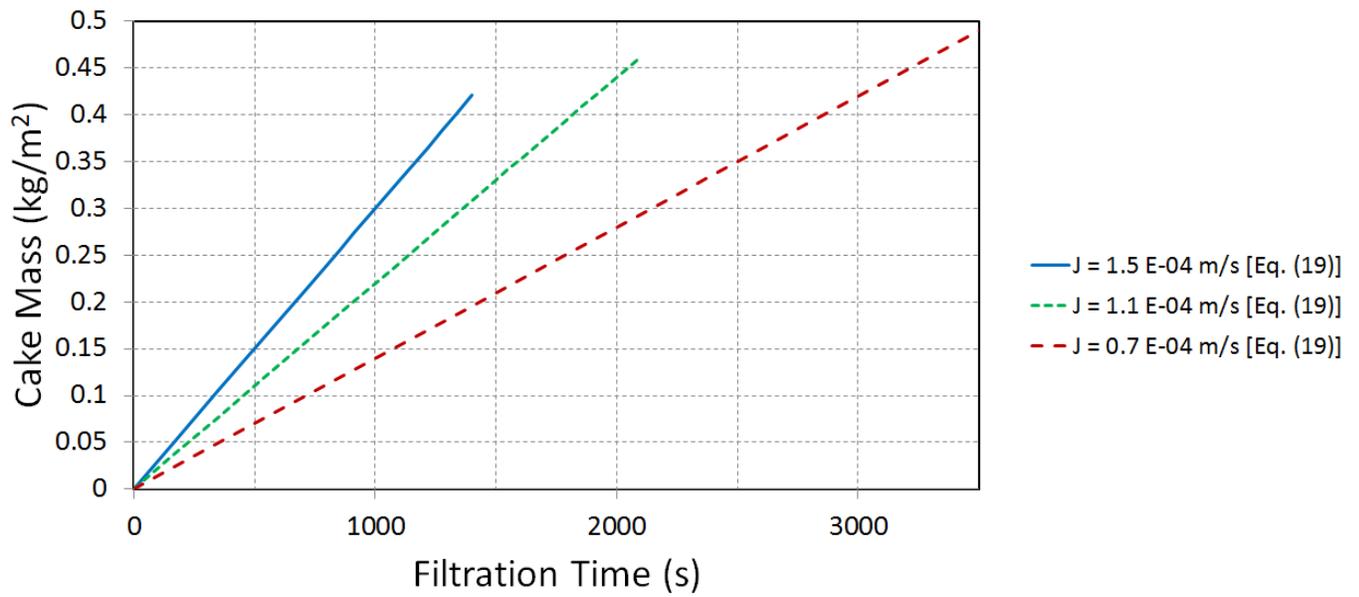

FIGURE 6 Predicted Cake Buildup [Eq. (19)] for the Experiments of Ho and Zydney[10] in the Microfiltration of a BSA Solution with a PCTE Membrane. Values of Model Parameters are provided in Table 5. Feed Concentration = 2 kg/m$^3$.

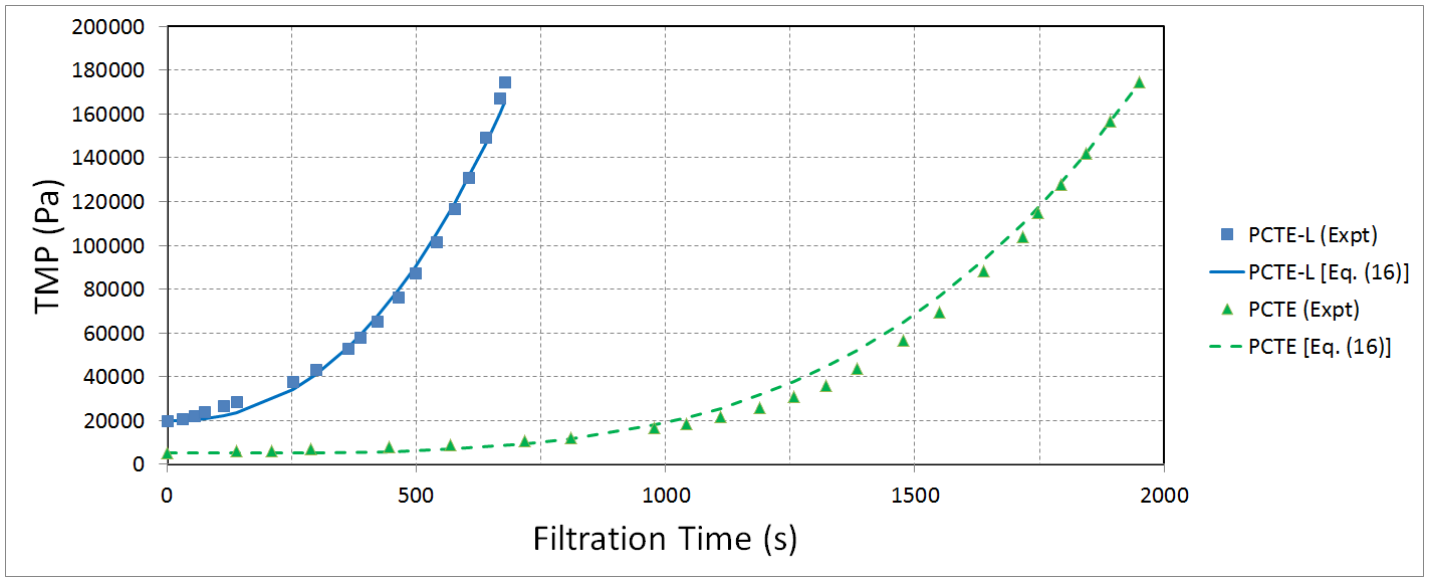

FIGURE 7 Comparison of the Complete Model [Eq. (16)] and Experimental (Fig. 2 of Ho and Zydney[10]) TMP Profiles in the Microfiltration of a BSA Solution with PCTE-L and PCTE Membranes. Values of Model Parameters are provided in Table 6. Feed Concentration = 2 kg/m$^3$ and Permeate Flux = 1.3 × 10$^{-4}$ m/s.

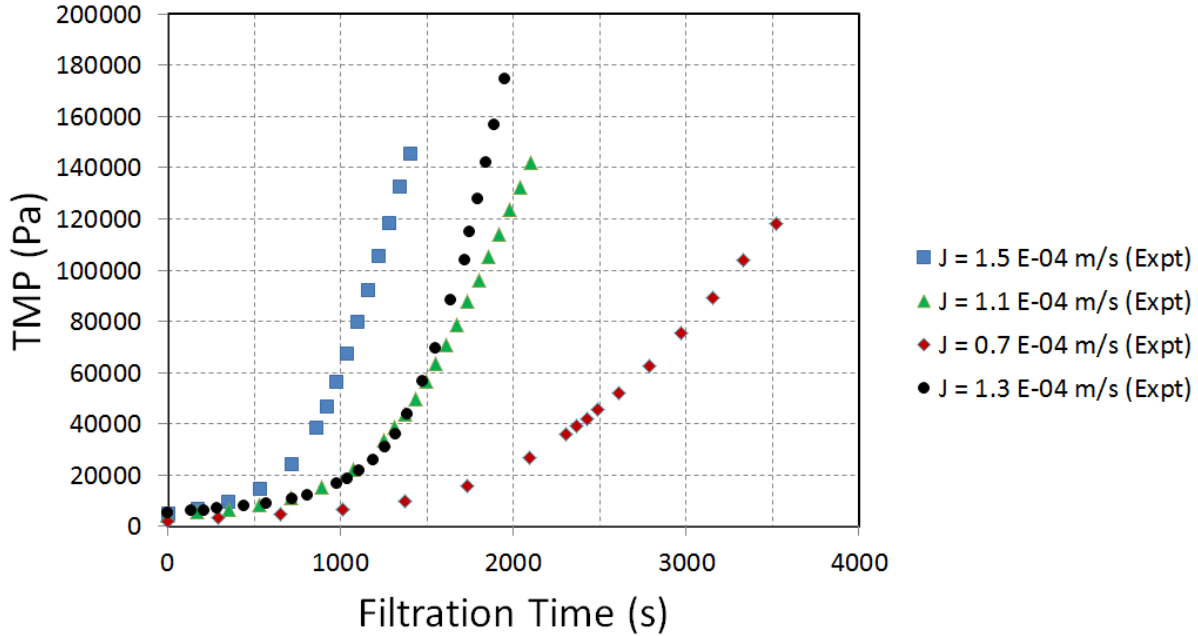

FIGURE 8 Experimental TMP Profiles (Figs. 1 and 2 of Ho and Zydney[10]) in the Microfiltration of a BSA Solution with PCTE Membranes. Feed Concentration = 2 kg/m$^3$.

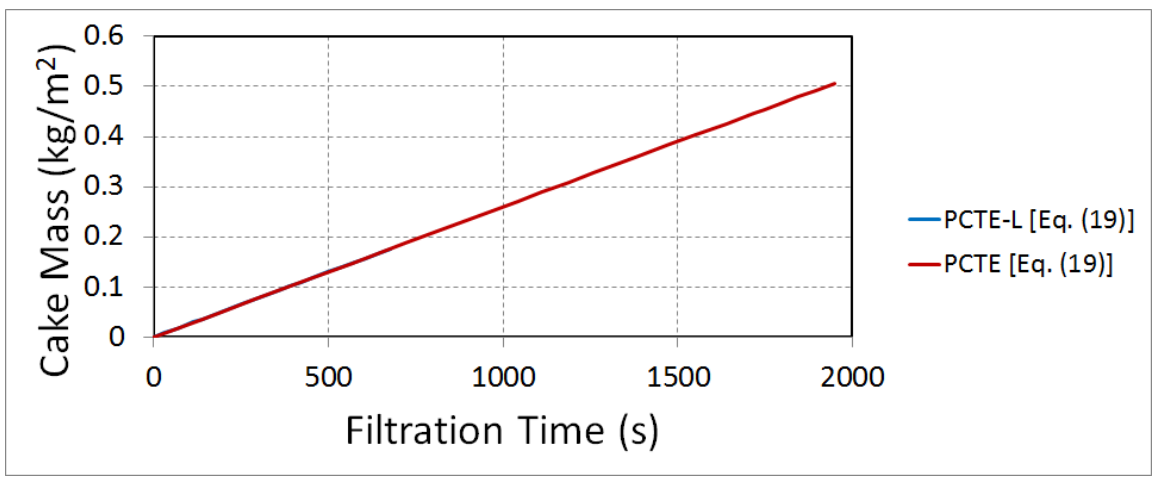

FIGURE 9 Predicted Cake Buildup [Eq. (19)] for the Experiments of Ho and Zydney[10] in the Microfiltration of a BSA Solution with PCTE-L and PCTE Membranes. Values of Model Parameters are provided in Table 6. Feed Concentration = 2 kg/m$^3$ and Permeate Flux = $1.3 \times 10^{-4}$ m/s.

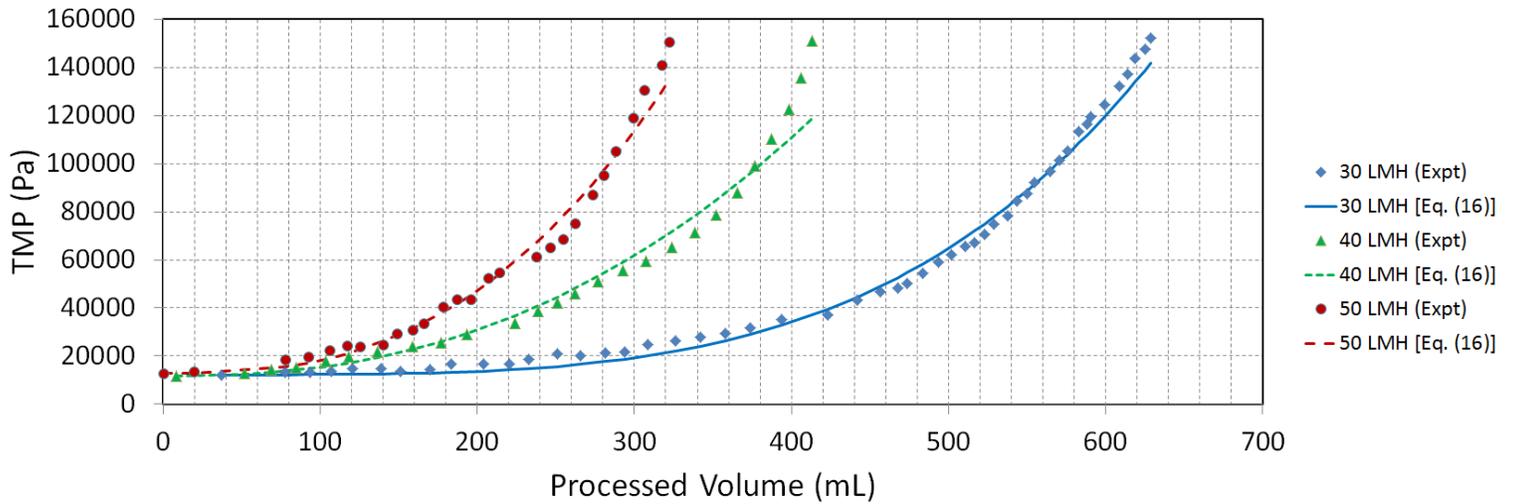

FIGURE 10 Comparison of the Complete Model [Eq. (16)] and Experimental (Fig. 5 of Kovalsky et al.[11]) TMP Profiles in the Microfiltration of a Yeast Suspension at pH 2.7. Values of Model Parameters are provided in Table 7. Feed Concentration = 10 kg/m$^3$ and LMH = L/m$^2$/h.

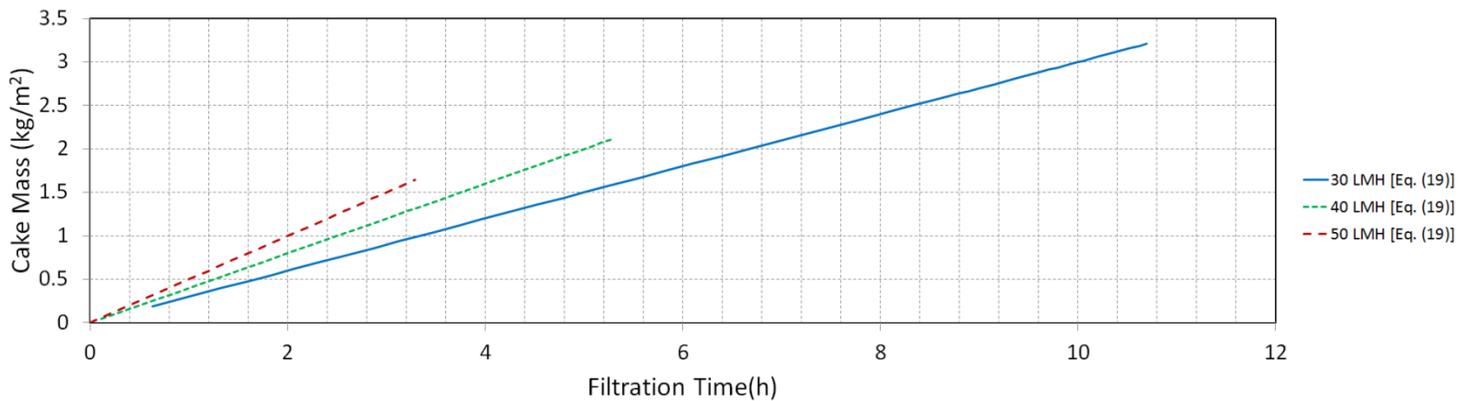

FIGURE 11 Predicted Cake Buildup [Eq. (19)] for the Experiments of Kovalsky et al.[11] in the Microfiltration of a Yeast Suspension at pH 2.7. Values of Model Parameters are provided in Table 7. Feed Concentration = 10 kg/m$^3$ and LMH = L/m$^2$/h.